\documentclass{optica-article}

\journal{opticajournal} 

\articletype{Research Article}

\begin{document}

\title{Color-switching in an optical parametric oscillator using a phase-conjugate mirror}

\author{B. E. Anderson,\authormark{1,*} J. Zhao,\authormark{2} Z. Zhou,\authormark{3} R. Speirs,\authormark{3} K. M. Jones,\authormark{4} and P. D. Lett\authormark{3,5}}

\address{\authormark{1}Department of Physics, The American University, Washington, DC, USA\\
\authormark{2}Centre for Quantum Computation and Communication Technology, Department of Quantum Science and Technology, The Australian National University, Canberra, ACT, Australia\\
\authormark{3}Joint Quantum Institute, University of Maryland, College Park, MD, USA\\
\authormark{4}Department of Physics, Williams College, Williamstown, MA, USA\\
\authormark{5}National Institute of Standards and Technology, Gaithersburg, MD, USA}

\email{\authormark{*}banderso@american.edu} 

\begin{abstract*} 
 We construct a phase-conjugate resonator which passively produces stable pulses that alternate between the probe and the conjugate colors. The requisite phase-conjugate mirror inside the resonator is constructed using non-degenerate four-wave mixing (4WM) in rubidium vapor. The glancing-angle phase-conjugate mirror is a 100\% output coupler, and therefore this resonator is unusual in that no light circulates the cavity more than once. Without the gain of the phase-conjugate mirror, the cavity boundary conditions, and thus resonant modes, are not defined and therefore can be tuned by the pump. The output of the optical parametric oscillator that is formed above threshold can passively mode-lock. The phase-conjugate mirror removes thermal or acoustic instabilities that are on a MHz or slower timescale.  This work provides a new method for stable pulsing using phase-conjugate optics, and suggests a platform for producing mode-locked pulses with squeezed light, as the 4WM process has already demonstrated quantum correlations.

\end{abstract*}

\section{Introduction}\label{sec1}

\begin{figure}[h]%
\centering
\includegraphics[width=\textwidth]{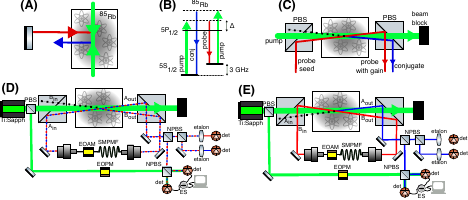}
\caption{The experimental setup. (A) A non-degenerate phase conjugate resonator. Two counter-propagating pump beams (green) optically pump an atomic vapor of $^{85}$Rb. An incoming red probe, detuned $-\Delta$ from the pumps, produces a blue non-degenerate phase conjugate, with detuning $+\Delta$. (B) The atomic level diagram for $^{85}$Rb with our 4WM scheme. The probe lies $\approx$3 GHz to the red of the pump and the conjugate is the same detuning to the blue. (C) Conventional single-pass forward 4WM. A single strong pump (green) beam optically pumps an atomic ensemble in the forward direction and is discarded at the black beam block. A probe seed is amplified and produces a non-degenerate blue phase conjugate at a glancing angle determined by phase-matching conditions. The black dashed line depicts the unused (vacuum) input port. The probe seed and pump have perpendicular polarizations and are combined on a polarizing beam splitter (PBS). (D) Our PCR. The output $B_{out}$ is recycled back to the input $A_{in}$ via a single-mode polarization maintaining optical fiber (SMPMF) causing the system to lase and produce (at $A_{out}$) pulses alternating between the red and blue modes. The SMPMF has a length ranging from several meters to over 100m. An etalon can be placed in $A_{out}$ to separate the red (probe) and blue (conjugate) pulses which are separated by 6 GHz in frequency. We can also choose to detect the light via optical heterodyne on a balanced detector. EOAM, electro-optic amplitude modulator; EOPM, electro-optic phase modulator; NPBS, non-polarizing beam splitter; ES, electronic subtraction; det, amplified photodiode detector. See text for further details. (E) Our CR-OPO. The output $B_{out}$ is recycled via fiber to $B_{in}$, leaving $A_{in}$ to be vacuum. The output at $A_{out}$ is a single color.}
\label{fig1}
\end{figure}

 Optical phase conjugation (OPC) allows a field to be generated that is effectively the time-reversal of an incoming wave \cite{fisher_optical_1983, boyd_nonlinear_2020}. Optical phase conjugation can be realized by non-degenerate four-wave-mixing (4WM) using a nonlinear medium with a significant third-order electric susceptibility. The medium that enables OPC acts as a phase-conjugate mirror (PCM), as shown in Fig. \ref{fig1}A \cite{feinberg_phase-conjugating_1980, yariv_phase_1978, liao_cw_1978, hellwarth_generation_1977, cronin-golomb_nondegenerate_1985, yariv_amplified_1977}. A notable application of PCMs is to reverse phase distortions in optical systems. The phase-healing properties of OPC have long been viewed as a mechanism that could be used to construct stable resonators with aberrated beams, and essentially self-healing resonators that can compensate for changing thermal or acoustic phase distortions \cite{bloom_conjugate_1977, beldyugin_properties_1979, pepper_compensation_1980, lind_demonstration_1981}. Such resonators have also been shown to be stable in longitudinal mode operation \cite{cronin-golomb_nondegenerate_1985, lam_optical_1979, auyeung_theoretical_1979}.

When phase conjugate optics were first developed in the late 1970s, a number of spectacular applications were foreseen, however most have not been realized or are realized only in demonstration \cite{lanzerotti_theory_1995}. Phase conjugate materials have typically been either slow to respond, have low gain, or are simply difficult to use \cite{nilsen_narrow-band_1981}. Although significant work has been done, several theoretical predictions for resonators based on PCMs have never been explored because of this.

Although the emphasis in OPC research has often been on correcting spatial distortions, here we explore a different application of PCM technology by constructing a type of optical parametric oscillator (OPO) called a phase-conjugate resonator (PCR) for stable mode-locking \cite{belanger_resonant_1980}. Although mode-locking using a PCM has been explored theoretically before \cite{gray_mode_1995}, and demonstrated experimentally \cite{vanherzeele_mode-locked_1981}, not much further work has been pursued.  Color switching behavior between frequency modes of a PCR was suggested in \cite{belanger_resonant_1980}, but has not been demonstrated experimentally. Conventionally, a PCR is constructed by replacing a mirror of an optical cavity with a PCM, as in Fig. \ref{fig1}A, where the PCM is an optically pumped gas of atoms. Here, as described in detail in the next section, we construct a PCR (Fig. \ref{fig1}D) from a glancing-angle PCM using non-degenerate 4WM in a forward phase-matched geometry and a ring cavity built from a fiber loop. Our aim is to investigate the longitudinal mode properties of this system.

This resonator is unusual in that reproducing an incoming wavefront requires two trips around its ring cavity. Therefore, the cavity mode spacing is halved as compared to an OPO that uses a conventional resonator (CR-OPO), as shown in Fig. \ref{fig1}E. We find that the PCR builds up optical power from spontaneous emission and, for a sufficiently long cavity, produces mode-locked pulses in the two non-degenerate colors supported by the 4WM. Although the gain bandwidth of our resonator is only tens of MHz, our resonator uses a fiber loop that can be made long enough to support many longitudinal modes. The resonator is also unusual in that there are no empty cavity modes in the absence of the gain in the 4WM medium, and even above threshold no light circulates more than once through the cavity.

\section{Results}
\subsection{Experimental set-up}
Our basic 4WM scheme is shown in Fig. \ref{fig1}B,C. The gain region can be viewed as a glancing-angle PCM using non-degenerate 4WM \cite{mccormick_strong_2007}. Our 4WM medium is a vapor of $^{85}$Rb at $100\:^{\circ}$C contained in a 12 mm long glass cell. The atoms are pumped by a continuous-wave Ti:sapphire laser producing a 500 mW, 1 mm diameter beam tuned blue of the D1 (795 nm, 5S$_{1/2}$ to 5P$_{1/2}$) transition by $\Delta = 800$ MHz. This detuning is chosen empirically to optimize the output power by finding a compromise between absorption in the atomic vapor and non-linear gain. The 4WM cell acts as an optical amplifier with two input ports and two output ports. If an independent probe beam detuned $\approx$3~GHz (the ground state splitting in $^{85}$Rb) to the red of the pump crosses the pump beam at an angle of approximately $0.5^{\circ}$ inside the cell as in Fig \ref{fig1}C, it is amplified via the 4WM. The 4WM also produces a phase conjugate beam $\approx$3~GHz to the blue of the pump, which exits at the same small angle to the pump on the opposite side of the pump beam from the probe. The angle between the pump and incident beam is chosen to facilitate phase-matching in the atomic medium. Since the output probe and conjugate are slightly different wavelengths from the pump, they must exit at an angle such that $2\vec{k}_{\text{pump}} = \vec{k}_{\text{probe}} + \vec{k}_{\text{conj}}$. We note that this angle is determined by the index of refraction of the atomic vapor, which is due to the vapor's significant dispersion. In the experiment we tend to find this angle empirically by aligning an input probe seed relative to the pump until we see a conjugate stimulated by the four-wave mixing. A typical single-pass small-signal gain observed under these circumstances is of order 10. This 4WM configuration is often called `nearly nondegenerate' as the detuning of the probe or conjugate from the frequency of the pump is small compared to the pump frequency itself \cite{fisher_optical_1983}.

We can create a resonator in two distinct ways. First, in Fig. \ref{fig1}D, the signal out at $B_{out}$ is sent back into the input port $A_{in}$ (a PCR). Note that when probe frequency light is sent into $A_{in}$, conjugate frequency light emerges from $B_{out}$ and vice versa. In the second resonator construction, in Fig. \ref{fig1}E, the signal at $B_{out}$ is sent back into the input port $B_{in}$ (a CR-OPO). In this case, whichever frequency light is input at $B_{in}$, the same frequency light is produced at $B_{out}$. In either case of the PCR or CR-OPO, there is no input seed in normal operation. Before the cavity is properly aligned with beam angles commensurate for phase-matching, and without an independent input probe, the system will not build up optical power. However, when the output of an optical fiber is aligned to either $A_{in}$ or $B_{in}$, with an angle dictated by phase matching in the 4WM, optical power can build up on its own from spontaneous emission. 

With the fiber installed as in Fig \ref{fig1}D, the maximum output power of our system is approximately 30 mW. Stable lasing can be achieved with significantly lower pump powers, however this conversion efficiency of approximately 6\% provided ample power for the simultaneous measurements shown in Fig. \ref{fig1}D,E. This conversion efficiency is primarily limited by the insertion loss of the optical fiber from free space, fiber-to-fiber connectors, and long optical fiber lengths. The length of optical fiber used is varied between several meters and 105 m in length. Because the 4WM is sensitive to polarization, only single-mode polarization-maintaining optical fibers (SMPMF) are used. The PCM is a $100\%$ output coupler, and therefore any light entering the ring cavity leaves with gain when it returns to the PCM after one round trip \cite{fisher_optical_1983}. The phase-conjugate mirror essentially provides a ``copy" of the incoming beam, in the form of a phase-conjugate, that is then injected back into the cavity to circulate the cavity once. When the phase-conjugate arrives at the vapor cell one round trip time later, it immediately exits, but not before stimulating a phase-conjugate of itself which is then sent into the cavity. This cycle repeats. In a conventional resonator, the cavity defines resonant modes on its own. In contrast, this PCR configuration contains no empty cavity modes without the gain medium to recirculate the light.

\begin{figure}[h]%
\centering
\includegraphics[width=\textwidth]{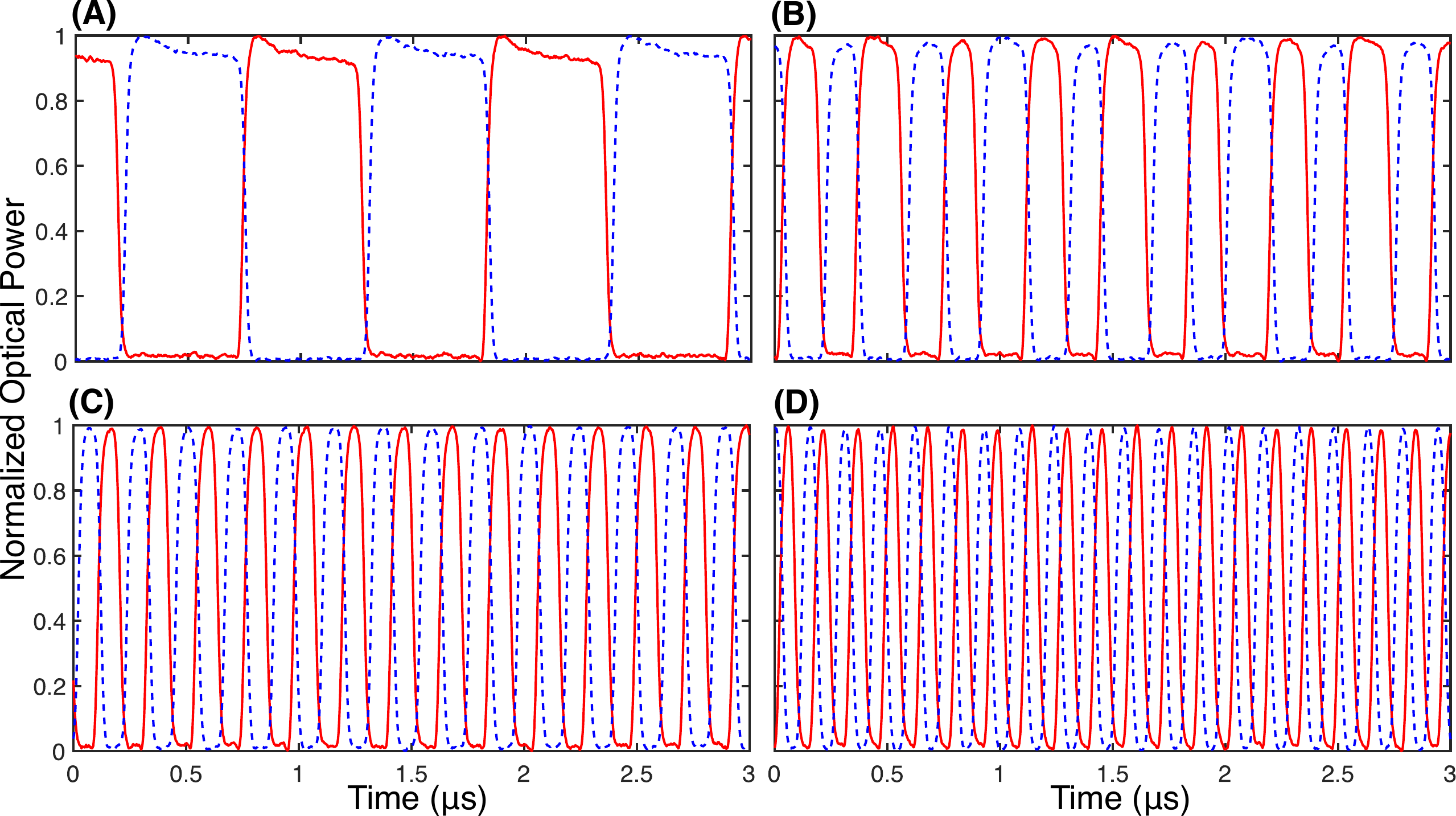}
\caption{Mode-locked pulses from the output of the PCR. The output $A_{out}$ in Fig. \ref{fig1}D is split into two and filtered with etalons to separate the two colors of light. The probe is shown in solid red and the conjugate is shown in dotted blue. The optical power is normalized to one for both colors. The bottoms of the pulses reach zero consistent with what is allowed by the 20~dB extinction ratio of the etalons. All four frames, (A)-(D), are produced under nominally the same experimental conditions and selected from a series of runs in which the laser is restarted each time from noise. Although the most likely pulse period is the one shown in (A), slight differences in the initial conditions lead to the system occasionally pulsing at higher frequencies as shown in B-D. The fiber length used is 105 m, corresponding to a 0.5 $\mu$s round trip time.}
\label{fig2}
\end{figure}

We measure the output $A_{out}$ which is separated into three paths via beam splitters. Etalons are used in the first two paths to separate the probe and conjugate light before sending them to detectors. The last path sends the light to optical heterodyne detection which is described below. The etalons have a bandwidth of approximately 1 GHz and each is centered on either the probe or conjugate carrier frequencies. Fig. \ref{fig2}A shows separately the probe and conjugate signals from $A_{out}$ using a 105 m optical fiber length, which corresponds to approximately a 0.5 $\mu$s round trip time. The output is mode-locked and consists of nearly square pulses that alternate between the probe and conjugate modes. Because the probe and conjugate frequencies are only separated by 6 GHz, if $A_{out}$ is measured on a standard near-infra-red optical detector without an optical filter we would not be able to discriminate between the probe and conjugate colors. Typically, the probe and conjugate pulses are of similar intensity, but either may be larger depending on the alignment of the cavity. The pulsing repeats itself after two round trip times.

\begin{figure}[h]%
\centering
\includegraphics[width=\textwidth]{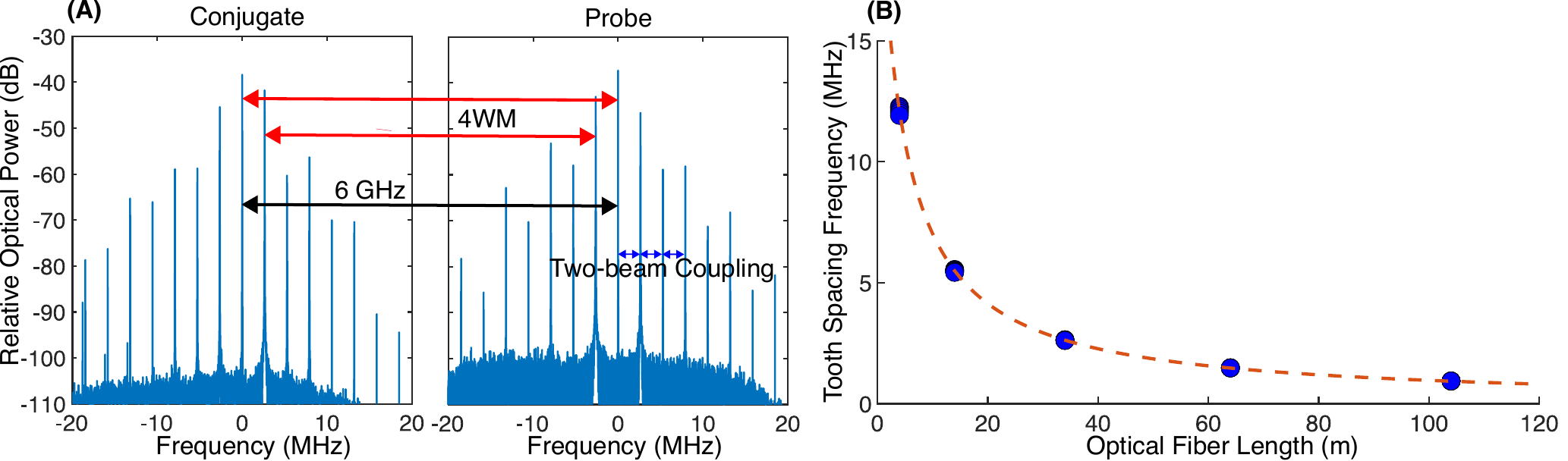}
\caption{Heterodyne spectra and comb frequency spacing of the PCR. (A) Heterodyne spectra of the probe and conjugate light from $A_{out}$ using a 34 m fiber. The centers of the combs are separated by 6 GHz and the pump is directly in the middle. Comb teeth of the probe and conjugate that are equally spaced from the pump are linked via 4WM (red arrows show two examples). Two-beam coupling links adjacent teeth (blue arrows show three examples) on the probe beam. This process does not involve the pump beam. The frequency centers of the spectra have been shifted to zero in post-processing for readability, and are double-sided spectra. (B) The tooth spacing frequency, measured in heterodyne data such as Fig. \ref{fig3}A, is the inverse of the pulse period. The fit shown is $\delta f=\frac{c}{2\left(1.5\times L_f+6.4\left[m\right]\right)}$ where $L_f$ is the optical fiber length. The factor of two comes from the PCR reproducing the wavefront after two roundtrips (one probe and one conjugate successively) around the cavity.}
\label{fig3}
\end{figure}

We can investigate the longitudinal mode structure of the field at $A_{out}$ via optical heterodyne measurements instead of direct measurement on a single photodiode. $A_{out}$ can be mixed in a non-polarizing beamsplitter with light from the pump that has $\pm 3$ GHz sidebands applied to it via an electro-optic phase modulator. Each of the sidebands will be close in frequency to the probe or conjugate such that the heterodyne beat frequency is small enough to be measured on an amplified photodiode. An example heterodyne signal is shown in Fig. \ref{fig3}A. The frequency spacing between comb teeth as a function of fiber length is shown in Fig. \ref{fig3}B. The data is well fit by the function $\delta f=\frac{c}{2\left(n_fL_f+L\right)}$, where $c$ is the speed of light, $n_f$ is the index of refraction of the optical fiber, $L_f$  is the length of the optical fiber, and $L$ is the effective remaining path length of free space and the atoms. The factor of 2 in the denominator reflects the assumption that two round trips of the cavity are required for the wavefront to be reproduced. The good agreement with the data confirms this expectation. Although the pulsing frequency of the PCR is set by the length of fiber, the maximum pulsing frequency in our PCR is set by properties of the atomic medium. The bandwidth of the 4WM process is limited by the natural linewidth of the excited state in rubidium which is on the order of 10 MHz, as discussed in Sec. \ref{sec_stability}. This bandwidth could potentially be significantly increased in other, likely solid-state, 4WM platforms.

\subsection{Understanding Pulsing in the PCR}
The 4WM process, in the absence of a cavity, is fundamentally spatially symmetric about the pump beam and permits simultaneous generation of both the probe and conjugate frequencies into the same spatial modes at low intensities. At high intensities, mode competition for the gain prevents both frequencies from being present simultaneously in the same spatial mode in the gain medium. Without some form of mode coupling an OPO would operate single-mode (or probe-conjugate mode-pair) above threshold \cite{fabre_cw_2000}.  An additional 4WM process known as 2-beam coupling \cite{boyd_nonlinear_2020,wu_two-beam_2021,kauranen_amplification_1994} provides a coupling between neighboring frequency modes at the probe color.  Thus, a set of frequency modes can stably oscillate together in our system.  The mode competition, however, forces the output to switch between probe and conjugate colors provided there is sufficient gain bandwidth to support multiple frequencies at each color, and thus pulse formation. As a high intensity probe beam exits the cavity at $A_{out}$, it stimulates the production of a high intensity conjugate into the resonator mode at $B_{out}$.

Although the resonator most often produces pulses with a comb tooth spacing of $\delta f=\frac{1}{2T}$, where $T$ is the round-trip time of the cavity, the resonator will sometimes output $n\,\delta\!f$ where $n$ is an odd natural number. In Fig. \ref{fig2}B-D we show tooth spacings $3\,\delta\!f$, $5\,\delta\!f$, and $7\,\delta\!f$. Surprisingly, this data is taken using identical experimental conditions as in Fig. \ref{fig2}A. To take the different datasets, we `re-start' the cavity by rapidly increasing and decreasing the optical loss of the cavity using a fiber electro-optic amplitude modulator placed in the path of the SMPMF. Empirically, we find the likelihood of seeing $n\,\delta\!f$ decreases as $n$ increases, and $7\,\delta\!f$ is only seen a few percent of the time. 

We can understand how these higher pulse frequencies are physically permitted by first considering the cavity operating at spacing $\delta\!f$ and just starting to output probe light. We consider the cavity to be filled with probe light which is on its way out of $A_{out}$. The probe then stimulates conjugate light to enter the cavity, filling in behind the probe light. After time $T$, the cavity is filled with conjugate light and it begins to exit $A_{out}$. This stimulates probe light to enter the cavity, filling in behind the conjugate light and completing the cycle. Another stable pattern occurs if the cavity is initially filled with three pulses of light: the first $1/3$ probe, the second $1/3$ conjugate, and the final $1/3$ probe. At time $T/3$ later, the first probe light in the cavity will have exited, while stimulating conjugate light to fill in behind the final probe light. Now the cavity is filled $1/3$ by conjugate, $1/3$ by probe, and $1/3$ by conjugate. Another time $T/3$ later will reproduce the initial conditions.

If the resonator could operate at $2\,\delta\!f$, then the cavity would at some moment be filled $1/2$ by probe and $1/2$ by conjugate light. As the probe begins to exit, it stimulates conjugate light to fill the cavity. But this conjugate light is filling in behind the already existing conjugate light in the first half of the cavity. At time $T/2$ later, the entire cavity is filled with conjugate light and the system reverts to operating at mode spacing $\delta\!f$. Even multiples of $\delta\!f$ are not stable in the cavity, whereas odd multiples are stable. Random fluctuations in the initial turn-on from spontaneous emission choose from the available stable periods.

\subsection{Stability of CR-OPO versus PCR}\label{sec_stability}
The PCR is stable against acoustic and thermal noise because the OPC process has no net phase accumulation after two round trips. This contrasts with the instability of a CR-OPO. Acoustic and thermal changes in the long fiber lengths cause drifts in the path lengths which affect the pulse timing. The stability of the PCR is due to the speed of response of the phase-compensation of the 4WM, which is dictated by optical pumping times; on the order of 1 $\mu$s. As mentioned above, a homogeneously-broadened gain medium, such as a CR-OPO, should have strong mode competition due to gain saturation, and with a single-mode pump this should result in the CR-OPO operating on only a single mode if there is no coupling between the available modes. Mode coupling continuously feeds population to cavity longitudinal modes that would otherwise be suppressed by the mode competition, and therefore prevents a single mode from winning out. In both our CR-OPO and PCR, a mode coupling mechanism exists via two-beam coupling which feeds neighboring modes and allows the system to oscillate stably in many modes. This two-beam coupling is essentially another 4WM process that mixes adjacent probe-color modes that are separated by a small enough frequency difference, as in Fig. \ref{fig3}A.

Two-beam coupling is possible in our 4WM system as long as different modes are on the order of a natural linewidth (6 MHz) apart. Given our 3 dB gain bandwidth of about 20 MHz, if there are three modes that can compete in this range, they are already coupled even at low intensities in this system. The two-beam coupling is self-pumped in that the pump beam driving the high-gain 4WM process indicated in Fig. \ref{fig1}B is not involved; only the neighboring probe-colored cavity modes mix and exchange energy as in Fig. \ref{fig3}A. The conjugate frequency is too far from resonance for the coupling to be significant at these intensities. 

As the cavity length is made shorter, the mode-spacing increases and the number of modes that can lase is reduced. Eventually, with a short enough cavity, the mode spacing 
is much larger than the approximately 20 MHz bandwidth and only a single mode of each probe/conjugate color can lase. Both colors then lase simultaneously and continuously in the resonator. In this limit, the color-switching behavior of the PCR system is lost. The lasing is stable regardless of the cavity length, as the PCM allows lasing at any arbitrary cavity length. With a slightly longer cavity we see sinusoidal oscillations on a continuous-wave background due to the beating of the central mode with weak sidemodes that start to lase one cavity mode spacing away. These oscillations are approximately out of phase in the probe and conjugate colors, although on the short time scale set by a fiber of several meters, group velocity delays in the gain medium become significant, as will be discussed in a later publication.

\begin{figure}[h]%
\centering
\includegraphics[width=\textwidth]{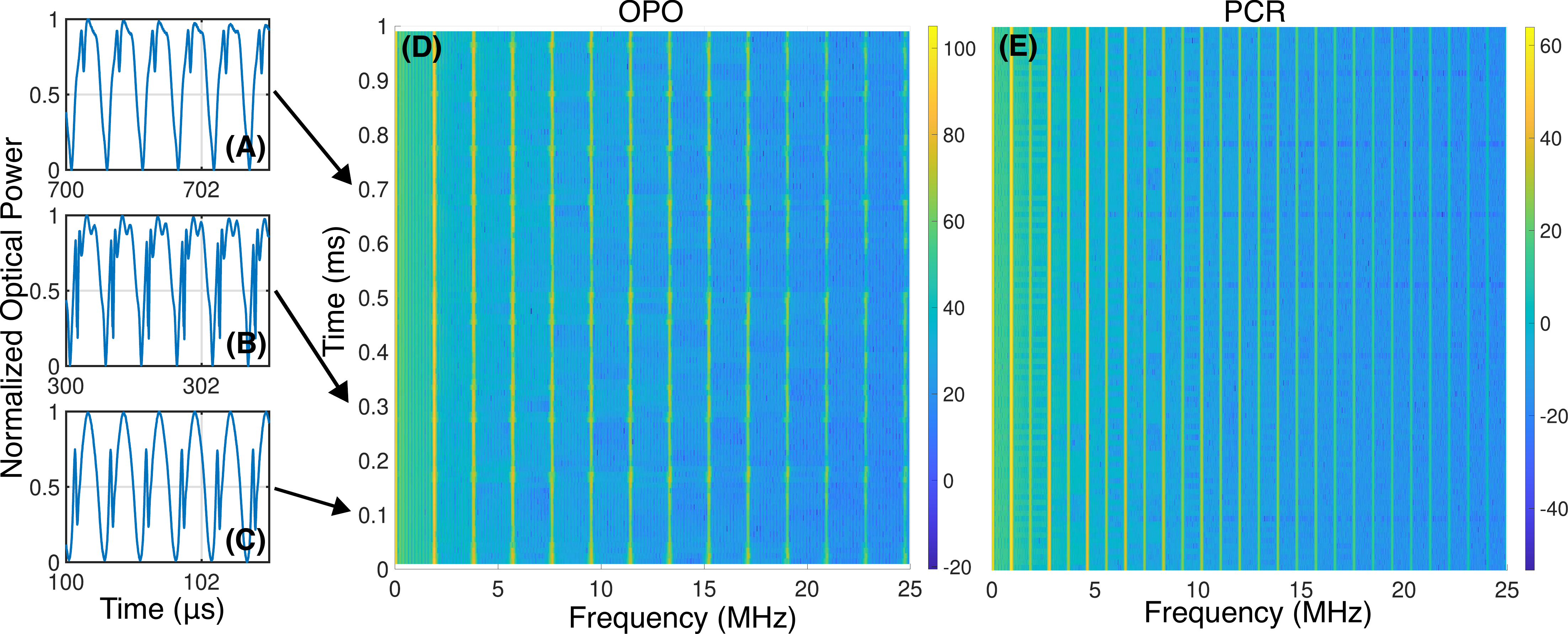}
\caption{Comparison of the outputs of the CR-OPO and PCR. (A-C) Output optical power of the CR-OPO at different times showing the changing waveform due to instability in the CR-OPO. (D) A spectrogram of the CR-OPO output showing regular instability on the order of 100 $\mu$s. The time resolution of the spectrogram is 20 $\mu$s, as explained in the text. The color scale for the z-axis shows optical power plotted in dB. (E) A spectrogram of the probe-color modes of the PCR with the same fiber length. The lowest mode is half the frequency of the CR-OPO and the lasing modes do not show the instability present in the CR-OPO, despite the PCR having no active stabilization.}
\label{fig4}
\end{figure}

\subsection{CR-OPO Configuration}
We can contrast the behavior of the system operated as a PCR (Fig. \ref{fig1}D) to that of the system operated as an CR-OPO (Fig. \ref{fig1}E). Figure \ref{fig4} shows data which allows us to identify three key differences in the output from the two configurations. First, the output of the CR-OPO is single color, either probe or conjugate. Second, the period of the CR-OPO output is half that of the PCR for a given optical path length. One traversal through the resonator constitutes a full period. Third, the output is significantly more unstable than the PCR. In Fig. \ref{fig4}A-C we can see the pulse shape of the CR-OPO evolves on the timescale of 100 $\mu$s. By taking spectrograms of the PCR and CR-OPO output, we can more easily compare their longer-timescale behavior. To do this, as shown in Fig. \ref{fig4}D, we record 1 ms of direct optical detection of $A_{out}$ from the CR-OPO, which has power only at the probe frequency. In post-processing, we then plot the power spectrum of the time-domain signals, with example time-domain data given in Fig \ref{fig4}A-C, using successive 20~$\mu$s windows as a function of time. These power spectrum windows are assembled together to make a spectrogram. A similar measurement is done for the PCR in Fig. \ref{fig4}E using the only the output of the probe etalon since the PCR produces both probe and conjugate light. Both figures use a 105 m fiber length, and we can see that the minimum frequency mode spacing of the PCR is half that of the CR-OPO. Instability in the CR-OPO manifests as blurry wider areas on the otherwise regular modes that are due to optical path length drift. We can contrast this to the PCR with its stable output.  

\begin{figure}[h]%
\centering
\includegraphics[width=\textwidth]{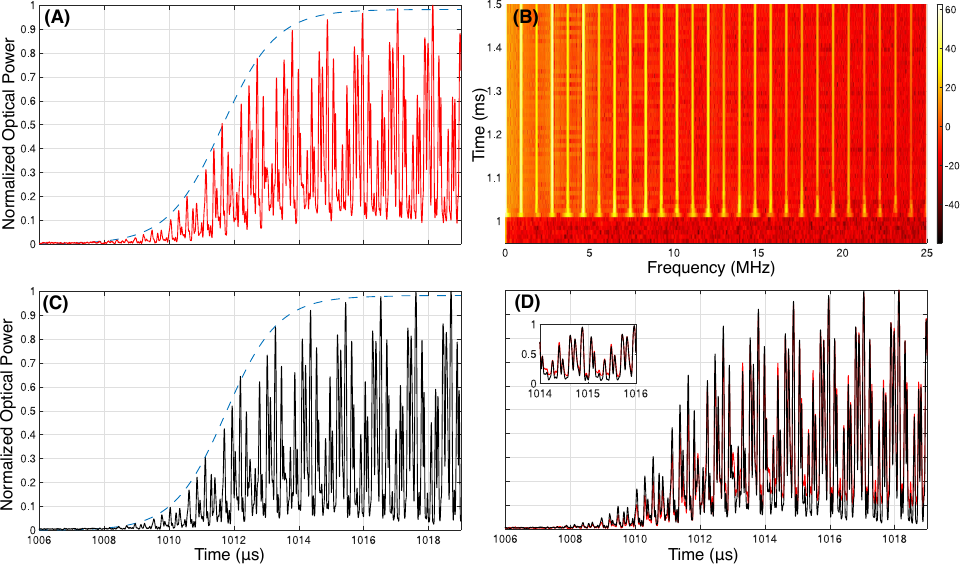}
\caption{The turn-on dynamics of the PCR with a 105 m fiber. (A) The probe output of the PCR. The PCR is rapidly switched on via a fiber electro-optic amplitude modulator at time = 1~ms. The dashed line provides a logistic curve guide of the envelope for the eye. (B) A spectrogram of the probe light showing the quick convergence to regularly spaced modes. The color scale for the z-axis shows relative optical power plotted in dB. (C) The conjugate output of the PCR. (D) In post-processing, the conjugate (black) output is shifted in time by one round trip time of the resonator and overlaid on the probe (red) output. The outputs are normalized in amplitude.}
\label{fig5}
\end{figure}

\subsection{Turn-on Dynamics}
To understand the dynamics of the PCR when it begins to lase, we inserted an electro-optic amplitude modulator to rapidly vary the optical loss and thus bring the system from below the lasing threshold to above. Figure \ref{fig5}A (Fig. \ref{fig5}C) shows the output at the probe (conjugate) frequency after the loss is lowered at $t=1$ms. A spectrogram of the probe in Fig. \ref{fig5}B shows the quick convergence of the system to regularly spaced modes after turning on. While the PCM sets the same maximum gain phase for each mode at the mirror position, dispersion inside the cavity would cause the mode spacing to be non-uniform. The only significant dispersion in this system over the small gain bandwidth is caused by the gain in the atomic vapor/4WM medium itself.  Any non-uniformity in the mode spacing is quickly removed as typical Kerr mode-locking effects that balance the dispersion and nonlinearities in the coupled-mode system cause the formation of regular pulses. The complex dynamics shown in Fig. \ref{fig5}A,C stabilize after several hundred microseconds to the regular square-like pulses shown in Fig. \ref{fig2}. During the buildup of optical power in the resonator the probe and conjugate match each other well in shape. By shifting the data for the conjugate beam forward in time by one period $T$ in post-processing, one can see that the two signals match well, as shown in Fig. \ref{fig5}D.

\section{Conclusion}
Aspects of the behavior shown here are reminiscent of behavior seen in Kerr soliton systems \cite{herr_temporal_2014,ackerhalt_solitons_1986}. Our 4WM mixing process provides nonlinearity and gain, as well as dispersion. In our system dispersion in the fiber is negligible over our limited bandwidth. Previous work has demonstrated the dispersion of the non-linear process via a difference in group velocity between probe and conjugate pulses \cite{boyer_ultraslow_2007}. Absorption in the atomic medium and coupling loss into the optical fiber provide dissipation. These are the ingredients of soliton formation in dissipative Kerr frequency combs. In the current observations reported above, the pulse duration is primarily set by the cavity length rather than the properties of the medium, suggesting the pulses are not solitons. Possibly under different conditions, the PCR could be modified to produce solitons.

The PCR does not, by itself, define mode positions without gain.  The maximum gain position can be tuned by the detuning of the pump frequency. Thus, arbitrary mode positions can be obtained enabling the system to have one of the oscillating modes placed at any frequency within the gain bandwidth.  This freedom to place a lasing line at any arbitrary frequency often requires physically modifying the resonator in frequency comb devices.

We have established a method for producing mode-locked pulses that exhibit a novel color-switching behavior. In the PCR configuration, the system is naturally stable in longitudinal mode due to the phase-conjugate mirror stabilizing accumulated phase after two round trips of the cavity. The 4WM process used here has been shown to generate strong quantum correlations (2-mode squeezing between the probe and conjugate beams) under similar detuning conditions and with low enough loss \cite{mccormick_strong_2007,gaeta_quantum_1988,agarwal_quantum_1993}. A suitably optimized PCR system might provide outputs which are quantum correlated. 

This proof-of-principle demonstration shows that one can use a phase-conjugate resonator to  create stable, mode-locked, pulses between two frequencies of light. Although the bandwidth of the system is limited, the physics of this color-switching process is novel and it could potentially be utilized in larger bandwidth 4WM systems or parametric down conversion systems. By cavity dumping the PCR, it is conceivable that one could extract large power quantum-correlated twin pulses. 

\begin{backmatter}
\bmsection{Funding}
Support for KMJ’s contribution was provided by the National Science Foundation. This work was supported by the Air Force Office of Scientific Research grant FA9550-16-1-0423.

\bmsection{Acknowledgments}
We thank Kartik Srinivasan, Gregory Moille, and Jared Wahlstrand for useful discussions.

\bmsection{Disclosures}
The authors declare no conflicts of interest.

\bmsection{Data availability} Data underlying the results presented in this paper are not publicly available at this time but may be obtained from the authors upon reasonable request.

\end{backmatter}

\bibliography{zotero.bib}

\end{document}